\documentclass[pra,twocolumn,showpacs,preprintnumbers,amsmath,amssymb,superscriptaddress,unsortedaddress]{revtex4}

\usepackage{graphicx}
\usepackage{dcolumn}
\usepackage{bm}
\usepackage{color}

\begin{document}

\preprint{}

\title{Violation of Clauser-Horne-Shimony-Holt inequality for states resulting from entanglement swapping}

\author{Antoni W\'{o}jcik}

\affiliation{Faculty of Physics, Adam Mickiewicz University, ul. Umultowska 85, 61-614 Pozna\'{n}, Poland}

\author{Joanna Mod{\l}awska}

\affiliation{Faculty of Physics, Adam Mickiewicz University, ul. Umultowska 85, 61-614 Pozna\'{n}, Poland}

\author{Andrzej Grudka}

\affiliation{Faculty of Physics, Adam Mickiewicz University, ul. Umultowska 85, 61-614 Pozna\'{n}, Poland}

\affiliation{National Quantum Information Centre of Gda\'{n}sk, ul. W{\l}adys{\l}awa Andersa 27, 81-824 Sopot, Poland}

\author{Miko{\l}aj Czechlewski}

\affiliation{Faculty of Physics, Adam Mickiewicz University, ul. Umultowska 85, 61-614 Pozna\'{n}, Poland}

\date{\today}

\begin{abstract}
We consider violation of CHSH inequality for states before and after entanglement swapping. We present a pair of initial states which do not violate CHSH inequality however the final state violates CHSH inequality for some results of Bell measurement performed in order to swap entanglement.
\end{abstract}

\maketitle

It is well known that any state which violates some Bell inequality is entangled. However, the converse is not true -- there are states which are entangled but correlations between distant measurements can be explained in terms of local hidden variables and hence the states do not violate any Bell inequality \cite{Werner1989}. Surprisingly, Popescu \cite{Popescu} and Gisin \cite{Gisin} showed that if particles in such a state are first subjected to local measurements then for some results of the measurements the post-measurement state can violate Bell inequality. 

Let us now suppose that we have two pairs of particles - the first pair is shared by Alice and Bob and the second pair is shared by Bob and Charlie. Each pair is in entangled state which does not violate Bell inequality. Let the parties perform entanglement swapping, i.e., Bob measures his particles in the Bell basis and sends a result of his measurement to Charlie \cite{Zukowski1, Zukowski2, Pan}. Charlie performs unitary operation which depends on a result of the measurement. We answer the following question stated by Sen {\it et al.}  \cite{Sen2}: {\it Can the resulting state of Alice's and Charlie's particles violate Bell inequality at least for some results of Bob's measurement?} for very important class of Bell inequalities, namely Clauser-Horne-Shimony-Holt (CHSH) inequalities \cite{Clauser}. In particular we give an example of two initial states which do not violate any CHSH inequality, however, if the parties perform entanglement swapping, then the final state of Alice's and Charlie's particles violates CHSH inequality for two results of Bob's measurement. 

Before we proceed to our example we briefly describe what was done before. In Ref. \cite{Sen2} the authors considered entanglement swapping in various configurations with three and more parties. In standard three-partite scenario they found an example of initial states which {\it do} violate CHSH inequality by some factor and the final state violates CHSH inequality by a {\it greater} factor. Moreover, they showed that in more general configurations with at least three pairs of particles the final multiparticle state  (i.e., the state of at least three particles) resulting from {\it generalized} entanglement swapping violates Mermin-Klyshko inequalities \cite{Mermin, Ardehali, Belinskii, Roy, Gisin2}, even if each initial two-particle state does not violate Bell inequality.

Let us now present our example. As was stated before we consider three parties -- Alice, Bob and Charlie. Alice and Bob share a state
\begin{eqnarray}
\label{Eq:1}
\varrho(p, \alpha)_{AB}=(1-p)|\Psi(\alpha) \rangle \langle \Psi(\alpha)|_{AB}+p|00\rangle \langle 00|_{AB},
\end{eqnarray}
where
\begin{eqnarray}
\label{Eq:2}
|\Psi(\alpha) \rangle_{XY}=\sin\alpha|01\rangle_{XY}+\cos\alpha|10\rangle_{XY}.
\end{eqnarray}
Bob and Charlie share a state (note the difference in the last term)
\begin{eqnarray}
\label{Eq:3}
\varrho(p, \alpha)_{BC}=(1-p)|\Psi(\alpha) \rangle \langle \Psi(\alpha)|_{BC}+p|11\rangle \langle 11|_{BC}.
\end{eqnarray}
The above states were also used to demonstrate that one can increase the maximally entangled fraction in the process of entanglement swapping \cite{Modlawska3}.

Arbitrary two-qubit state violates CHSH inequality if and only if the violation parameter $r(\varrho)=\sqrt{\lambda_1+\lambda_2}$ is greater than $1$, where $\lambda_1$ and $\lambda_2$ are the two largest eigenvalues of the matrix $R(\varrho)^TR(\varrho)$. $R(\varrho)$ is defined by its matrix elements  $R(\varrho)_{ij}=\text{Tr}(\varrho \sigma_i \otimes \sigma_j)$, where $\{\sigma_i\}$ are Pauli matrices \cite{Horodecki8}.

For states in Eqs. \ref{Eq:1} and \ref{Eq:3} the eigenvalues of the matrix $R(\varrho)^TR(\varrho)$ are:
\begin{eqnarray}
\label{Eq:4}
& \lambda_1=(2p-1)^2, \nonumber\\
& \lambda_{2, 3}=((p-1)\sin(2\alpha))^2.
\end{eqnarray}
Substituting them into the definition of the violation parameter we find that it is given by
\begin{eqnarray}
\label{Eq:5}
r(p, \alpha)=(1-p)\sin(2\alpha)\sqrt{1+z(p, \alpha)},
\end{eqnarray}
where
\begin{eqnarray}
\label{Eq:6}
z(p, \alpha)=\max \{ 1,\left( \frac{2p-1}{(p-1)\sin(2\alpha)} \right)^2 \}.
\end{eqnarray}
Let us suppose that Bob performs the Bell measurement on his qubits, i.e., he performs entanglement swapping. If he obtains $|\Phi^\pm\rangle_{BB}$ as a result of the measurement, where
\begin{eqnarray}
\label{Eq:7}
|\Phi^\pm\rangle_{XY}=\frac{1}{\sqrt{2}}(|00\rangle_{XY}\pm |11\rangle_{XY}),
\end{eqnarray}
then Alice and Charlie share a state (after possible phase correction)
\begin{eqnarray}
\label{Eq:8}
& \varrho_{AC}=\frac{1}{p(|\Phi^\pm\rangle)}[(1-p)^2\cos^2\alpha\sin^2\alpha|\Phi^+\rangle\langle\Phi^+|_{AC}+\nonumber\\
& +p(1-p)\sin^2\alpha|01\rangle\langle01|_{AC}],
\end{eqnarray}
where $p(|\Phi^\pm\rangle)=(1-p)^2\cos^2\alpha\sin^2\alpha+p(1-p)\sin^2\alpha$ is probability that Bob obtains $|\Phi^\pm\rangle$ as a result of the measurement.

After little algebra we find that the eigenvalues of the matrix $R(\varrho)^TR(\varrho)$ for the state in Eq. \ref{Eq:8} are:
\begin{eqnarray}
\label{Eq:9}
\lambda_1=\left( \frac{1-\tilde{y}(p, \alpha)}{1+\tilde{y}(p, \alpha)} \right)^2, \nonumber\\
\lambda_{2, 3}=\left( \frac{1}{1+\tilde{y}(p, \alpha)} \right)^2,
\end{eqnarray}
where
\begin{eqnarray}
\label{Eq:10}
\tilde{y}(p, \alpha)=\frac{p}{(1-p)\cos^2\alpha}.
\end{eqnarray}
Proceeding as before we find that the violation parameter is given by
\begin{eqnarray}
\label{Eq:11}
\tilde{r}(p, \alpha)=\frac{1}{1+\tilde{y}(p, \alpha)}\sqrt{1+\tilde{z}(p,\alpha)},
\end{eqnarray}
where
\begin{eqnarray}
\label{Eq:12}
\tilde{z}(p,\alpha)=\max \{ 1, (1-\tilde{y}(p, \alpha))^2 \}.
\end{eqnarray}

From Eqs. \ref{Eq:5} and \ref{Eq:6} we obtain that the initial states do not violate CHSH inequality ($r \leq 1$) if
\begin{eqnarray}
\label{Eq:13}
& \sin^2(2\alpha) \leq \frac{4p}{1-p}, \nonumber\\
& \sin^2(2\alpha) \leq \frac{(2p-1)^2}{(p-1)^2}
\end{eqnarray}
and
\begin{eqnarray}
\label{Eq:14}
& \sin^2(2\alpha) \leq \frac{1}{2(1-p)^2}, \nonumber\\
& \sin^2(2\alpha) > \frac{(2p-1)^2}{(p-1)^2}.
\end{eqnarray}

From Eqs. \ref{Eq:11}, \ref{Eq:12}, and \ref{Eq:10} we obtain that the final state violates CHSH inequality ($r>1$) if
\begin{eqnarray}
\cos^2\alpha >\frac{p}{(1-p)(\sqrt{2}-1)} .
\end{eqnarray}
The last inequality is satisfied if $p<p^*$, where 
\begin{eqnarray}
p^*=\frac{\sqrt{2}-1}{\sqrt{2}}
\end{eqnarray}
and $\alpha < \alpha^*(p)$, where
\begin{eqnarray}
\cos^2\alpha^*(p) =\frac{p}{(1-p)(\sqrt{2}-1)} .
\end{eqnarray}

From Eqs. \ref{Eq:13} and \ref{Eq:14} we obtain that for $p\in(p',p^*)$, where $p'=\frac{1}{2}p^*(1+\sqrt{p^*})$, and $\alpha \leq \alpha^*(p)$ the initial states do not violate CHSH inequality.
 We also obtain that the initial states do not violate CHSH inequality for $p \leq p'$ and $\alpha \leq \alpha'(p)$, where
\begin{eqnarray}
\sin^2(2\alpha'(p)) = \frac{4p}{1-p} \text{ for }  p \leq \frac{1}{2}p^*
\end{eqnarray}
and
\begin{eqnarray}
\sin^2(2\alpha'(p)) = \frac{1}{2(1-p)^2}  \text{ for }  p \geq \frac{1}{2}p^*.
\end{eqnarray}

\begin{figure} 
\includegraphics[width=7.5truecm] {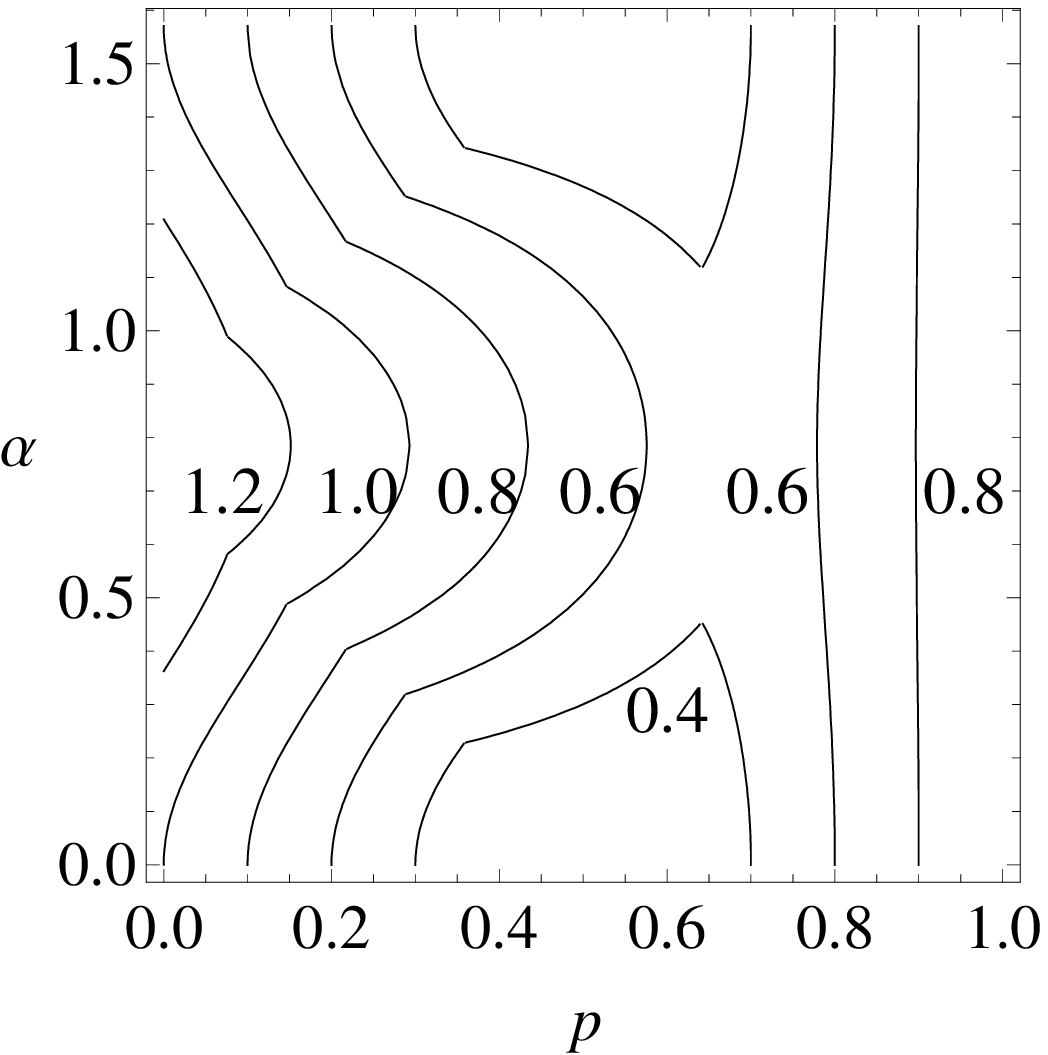}
\includegraphics[width=7.5truecm] {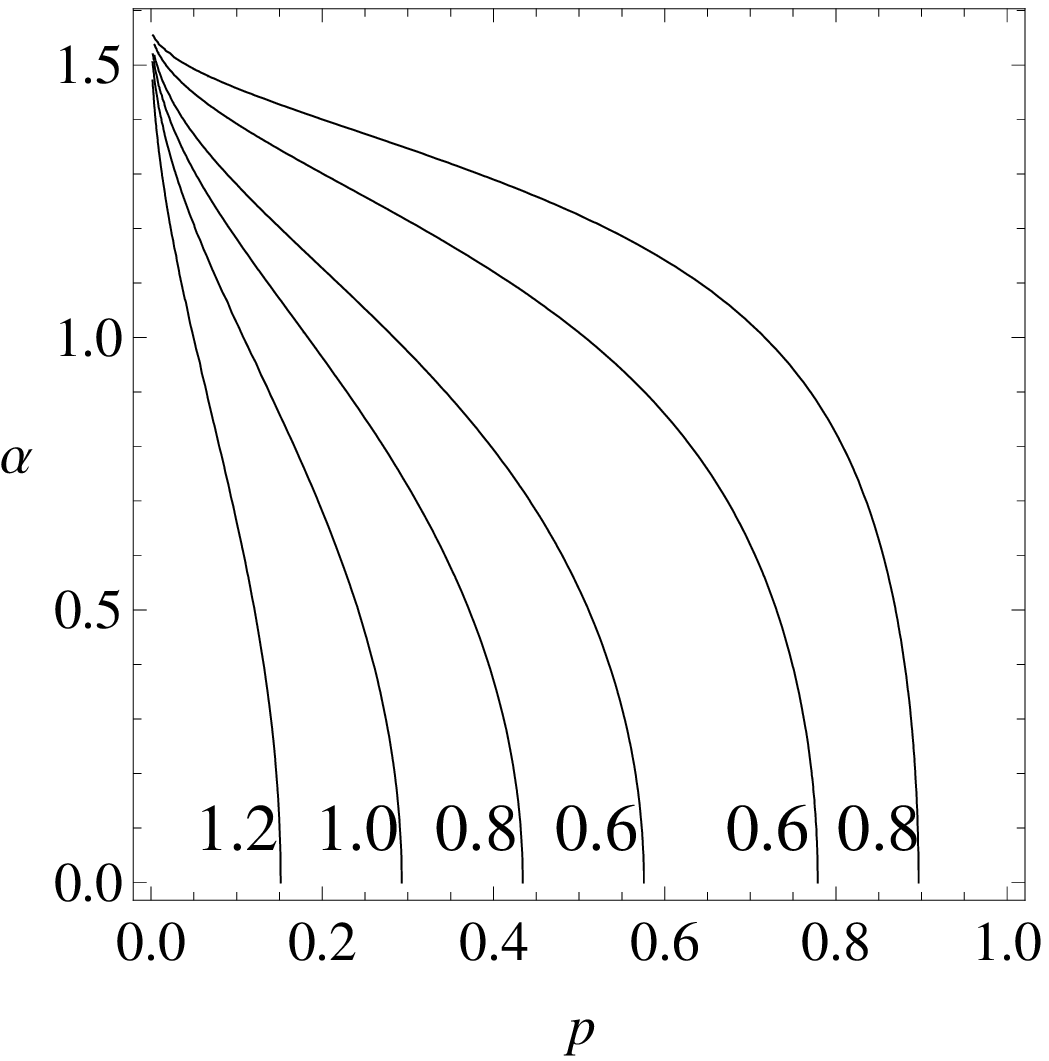}
\caption{\label{fig1}Dependence of the violation parameter $r$ on $\alpha$ and $p$ for the initial states (upper graph) and the final state (lower graph).} 
\end{figure}

In Fig. 1 we present dependence of the violation parameter $r$ on $\alpha$ and $p$ for the initial states and the final state. One can see that there exists a region where the states before entanglement swapping do not violate CHSH inequality and the state after entanglement swapping violates CHSH inequality.

As an example let us consider the initial states with $p=0.2$ and $\alpha \in[0,\pi/2]$ (see Fig. 2). On the one hand for $\alpha \leq 0.542051$ and $\alpha \geq 1.02875$ the initial states do not violate CHSH inequality. On the other hand for $\alpha < 0.681089$ the final state violates CHSH inequality. Hence for $\alpha < 0.542051$ the states before entanglement swapping do not violate CHSH inequality and the state after entanglement swapping violates CHSH inequality. 

\begin{figure} 
\includegraphics[width=7.5truecm]{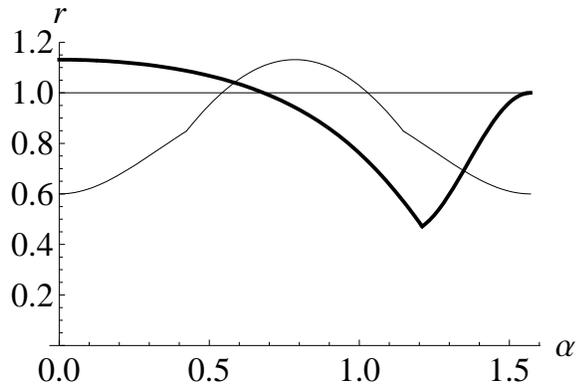}
\caption{\label{fig2}Dependence of the violation parameter $r$ on $\alpha$ for $p=0.2$ for the initial states (thin line) and the final state (thick line). Horizontal line denotes critical value of the violation parameter $r=1$.} 
\end{figure}

In conclusion we demonstrated what was called in Ref. \cite{Sen2} a kind of "superaddiditivity" in violation of CHSH inequality consequent on entanglement swapping. It is an open question if one can demonstrate "superadditivity" in violation of local realism because a state may satisfy CHSH inequality and violate some other Bell inequality \cite{Vertesi}.

\begin{acknowledgments}
We thank Antonio Acin, Nicolas Brunner and Marek \.{Z}ukowski for very helpful discussions. J.M. was partially supported by the Foundation for Polish Sciences.  A.G. was partially supported by Ministry of Science and Higher
Education Grant No. N N206 2701 33. 
\end{acknowledgments}

\end{document}